\documentclass[12pt]{article}
\usepackage{graphics,amssymb,latexsym,amsfonts,graphics}

\usepackage{amsmath, amssymb, graphics}

\begin{document}
\thispagestyle{empty}

\begin{center}
{\textbf{\Large{ Effect of temperature on the binding energy of a
 shallow hydrogenic impurity in a Quantum Well Wire }}}\\
T G Emam\\
 Department of Mathematics, Faculty of science, Ain Shams
university, Abbassia, Cairo, Egypt.\\
and\\
 Department of Mathematics, the German university in
Cairo-GUC, New Cairo city,  Main Entrance Al Tagamoa Al Khames,
Egypt.
\end{center}
\textbf{Abstract}\\ This work studies the effect of temperature on
the binding energy of a shallow hydrogenic impurity located on the
axis of a cylindrical semiconductor $GaAs Al_x Ga_{1-x} As$
quantum well wire. The results show that the binding energy at a
low temperature of $100 K$ is increased by about $11\%$ over that
associated with nearly room temperature, $300 K$, and is also
increased by
about $23\%$ over that associated with temperature of $500 K$.\\

\section{Introduction}
The electronic  properties of a hydrogenic impurity in the quantum
well[1,2,3,4] and the quantum well wire(QWW)[5,6,7,8,9,10] has
been the subject of many articles . The binding energy of a
shallow hydrogenic impurity in a quantum well wire has been
calculated as a function of the wire radius [5]. It has been found
that the binding energy increases with decreasing the wire radius
for an infinite well whereas it reaches a peak value and then
diminishes to a limiting value  for a finite potential. The effect
of location [6,7] of impurities has been studied as well as the
different shapes of the cross-section [6,7] of the wire. Chen et
al [11] have studied the effect of electron-phonon interaction on
the impurity binding energy in a quantum wire .\\
In this work, the effect of temperature on the  binding energy of
a donor impurity confined in a QWW is investigated theoretically
using a variational method .

\section{Theory}
The Hamiltonian for an on-axis hydrogenic impurity in a
cylindrical infinite  QWW of radius R is given by
\begin{equation}
H=\frac{-\hbar^2 \nabla^2}{2
m_{1,2}(T)}-\frac{e^2}{\epsilon_{1,2}(T)
\sqrt{\rho^2+z^2}}+V_b(T),
\end{equation}
where $m_{1,2}(T)$ and $ \epsilon_{1,2}(T)$ are the temperature
dependent effective-mass and dielectric constant of the QWW
respectively. The two subscripts $1,2$ stand for the potential
well (GaAs) and the potential barrier $(Al_x Ga_{1-x}As)$
materials respectively. The values of $m_{1,2}(T)$ and $
\epsilon_{1,2}(T)$ are taken as given in Refs.[12] and [13].  The
potential barrier height $V_b(T)$ is obtained from the temperature
dependent band-gap discontinuity $\Delta E_g(x,T)$ given in eV as
\begin{equation}
V_b (T)=\begin{cases} 0  , & \rho \leq R  \cr
 & \cr
0.60 \Delta E_g(x,T),& \rho > R ,
\end{cases}
\end{equation}
where
\begin{equation}
\Delta E_g(x,T)=x[1.247-(1.15\times10^{-4})T]; \ \ x<0.45.
\end{equation}
\hspace{1 cm} The aim is to find the binding energy of  the
impurity using variational method
\subsection{Electron Eigenstates}
Considering the absence of impurity and imposing the boundaray
condition ; Continuity of the wavefunction and its first
derivative at The QWW radius; the ground state wavefunction is
found to be [5]:
\begin{equation}
 \Psi(\rho)=\begin{cases} N J_0(r_{10} \rho), & \rho \leq R \cr
  \cr
 N J_0(r_{10} R)\frac{ K_0(b_{10} \rho)}{K_0(b_{10} R)}  , &
\rho
> R
 \end{cases}
\end{equation} where N is the normalization constant, and
\begin{equation}
r_{10}=\sqrt{\frac{2 m_1 E_{100}}{\hbar^2}},
\end{equation}
and
\begin{equation}
b_{10}=r_{10}\frac{J_1(r_{10} R) K_0 (r_{10} R)}{J_0(r_{10} R)
K_1(r_{10} R)},
\end{equation}
and the eigenenergy is
\begin{equation}
E_{100}=\frac{\hbar^2}{2 m_1}r_{10} ^2=V_b(T)-\frac{\hbar^2}{2
m_2}b_{10}^2.
\end{equation}
The continuity of the derivative of the wave function wavefunction
$\psi(\rho)$ at $\rho=R$ gives the equation
\begin{equation}
r_{10}(\frac{m_2(T)}{m_1(T)})J_1(r_{10}R)=b_{10} \frac{J_0(r_{10}
R)}{K_0(b_{10} R)} K_1(b_{10} R),
\end{equation}
using Equation(7),and putting the length in the bohr radius units
($a_B= \frac{\epsilon_0 \hbar^2}{m_e e^2}$)and the energy in the
Rydberg units ($R_b=\frac{e^2}{2 \epsilon_0 a_B}$),where $m_e$ is the electron effective mass in room temperature. \\
We get
\begin{equation}
r_{10}=\frac{J_0(r_{10} R) K_1(b_{10} R)}{J_1(r_{10} R)
K_0(b_{10}R)}\sqrt{V_b(T)\frac{m_1^2(T)}{m_2(T)
m_e}-\frac{r_{10}^2 m_1(T)}{m_2(T)}}
\end{equation}
the first root of this equation for a certain temperature and wire
radius gives the value of $r_{10}$, and hence the energy
eigenvalue $E_{100}$\\
\subsection{Impurity Eigenstates}
Now we consider an on-axis hydrogenic impurity, the
Schr\"{o}dinger equation is no longer separable and the trail wave
function is taken as
\begin{equation}
\Psi(\rho)=\begin{cases}N J_0(r_{10} \rho) exp(-\lambda
\sqrt{\rho^2+z^2}),& \rho \leq R \cr
  \cr
 N J_0(r_{10} R) \frac{ K_0(b_{10} \rho) exp(-\lambda
\sqrt{\rho^2+z^2})} {K_0(b_{10} R)}, & \rho
> R
 \end{cases}
\end{equation}

The normalization constant N is given by
\begin{equation}
N^{-2}=-2 \pi \frac{d}{d \lambda}(A+B)
\end{equation}
where
\begin{equation}
A=\int_{\rho=0}^{\rho=R} d \rho \rho J_0 ^2(r_{10} R)K_0(2 \lambda
 \rho)
\end{equation}
and
\begin{equation}
B=\frac{J_0 ^2(r_{10} R)}{K_0 ^2(b_{10}
R)}\int_{\rho=R}^{\rho=\infty} d \rho \rho K_0 ^2(b_{10} R)K_0(2
\lambda
 \rho).
\end{equation}
Calculating the Potential energy$<V>$ and the kinetic energy
$<K.E>$, it is found that
 \begin{equation}
<V>=-4 \pi e^2 N^2
(\frac{A}{\epsilon_1(T)}+\frac{B}{\epsilon_2(T)})-2 \pi N^2 V_b(T)
\frac{d B}{d \lambda},
 \end{equation}
and{\small{
\begin{eqnarray}
\nonumber
 <K.E>=&-& 2 \pi N^2 \lambda^2 (\frac{\hbar^2}{2
m_e})(\frac{m_e}{m_1(T)}\frac{d A}{d
\lambda}+\frac{m_e}{m_2(T)}\frac{d B}{d \lambda})\\
\nonumber
 &-&2 \pi N^2 (\frac{\hbar^2}{2 m_e})(\frac{m_e}{m_1(T)}
r_{10}^2\frac{d A}{d \lambda}-\frac{m_e}{m_2(T)}b_{10}^2\frac{d
B}{d \lambda})\\
 &+&2 \pi N^2 (\frac{\hbar^2}{2 m_e})(2 \lambda
R^2)[K_0(2 \lambda R) J_0^2(r_{10} R)
(\frac{m_e}{m_1(T)}-\frac{m_e}{m_2(T)})]
\end{eqnarray}
}}
 but
\begin{equation}
V_b(T)=\frac{\hbar^2 r_{10}^2}{2 m_1(T)}+\frac{\hbar^2 b_{10}^2}{2
m_2(T)},
\end{equation}
The total energy $E(R,T)=<T>+<K.E>$ can be obtained using
equations(14),(15),(16)together with equation (11)
\begin{eqnarray}
\nonumber
 E(R,T)&=&\frac{m_e}{m_1(T)}r_{10}^2+
\frac{\lambda^2(\frac{m_e}{m_1(T)} \frac{d A}{d
\lambda}+\frac{m_e}{m_2(T)}\frac{d B}{d \lambda})+4
(\frac{\epsilon_0}{\epsilon_1(T)}A+\frac{\epsilon_0}{\epsilon_2(T)}B)}{\frac{d
A}{d
\lambda}+\frac{d B}{d \lambda}}\\
&-& \frac{2 \lambda R^2
(\frac{m_e}{m_1(T)}-\frac{m_e}{m_2(T)})(K_0(2 \lambda R)
J_0^2(r_{10} R))}{\frac{d A}{d \lambda}+\frac{d B}{d \lambda}}
\end{eqnarray}
in the last equation we put the length in the bohr radius units
$a_B$ and the energy in  the
Rydberg units $R_b$. \\
 The binding energy $E_b(R,T)=\frac{\hbar^2 r_{10}^2}{2 m_e}$is
 then given by:
 \begin{eqnarray}
 \nonumber
E_b(R,T)=&-&\frac{\lambda^2(\frac{m_e}{m_1(T)} \frac{d A}{d
\lambda}+\frac{m_e}{m_2(T)}\frac{d B}{d \lambda})+4
(\frac{\epsilon_0}{\epsilon_1(T)}A+\frac{\epsilon_0}{\epsilon_2(T)}B)}{\frac{d
A}{d
\lambda}+\frac{d B}{d \lambda}}\\
&+& \frac{2 \lambda R^2
(\frac{m_e}{m_1(T)}-\frac{m_e}{m_2(T)})(K_0(2 \lambda R)
J_0^2(r_{10} R))}{\frac{d A}{d \lambda}+\frac{d B}{d \lambda}}
 \end{eqnarray}
 According to the variational method,  the right-hand side of the last equation is maximized
  with respect to
 $\lambda $ to obtain a lower bound of the binding energy.\\
 \section{Results and Discussions}
 Figure (1)  shows the dependence of the binding energy on
 the wire radius R for three different values of temperatures,$T=100K, T=300K$,and $T=500K$,
 at aluminium concentration $x=0.3$.
 At a given temperature the binding energy has a peak at a certain
 value of the wire radius , $R_{max}$, for this value the Coulomb
  interaction between the electron and the impurity is maximum due
  to the high localization of the wave function inside the well.
  For $R>R_{max}$ the binding energy decreases since the electron
  become less confined to the impurity due to the fact that the
  wave function becomes more spread out over the well. For
  $R<R_{max}$ the probability of finding the electron outside the
  well increases , the wave function penetrates through the
  surrounding barrier region, which leads to decrease the binding
  energy.  The results shown in figure (1) show also that the
  difference between the binding energy at a temperature of $100 K$
  and that corresponds to a temperature of $300 K$ for a QWW
  radius of $0.05 a_B$ is about $11\%$. While this difference is
  about $23\%$ in case of $T=100K$ and $T=500K$ .In fact,
  this difference decreases by increasing the wire radius. \\
  Figure (2) shows how the binding energy varies with the
  temperature for a certain value of the wire radius, $R=0.5 a_B$,
  at three values of aluminium concentration $x=0.15, x=0.30$ , and $x=0.45$,
  it can be concluded from figure (2) that the binding energy
  increases by decreasing the temperature and increasing the
  aluminium concentration, x. Increasing the value of x means increasing the value of
  the potential barrier, which in turns leads to a stronger confinement of the  electron inside
  the QWW, and consequently greater binding energy.

  \section{Conclusion}
   The effect of temperature on the binding energy of a shallow
   hydrogenic impurity in a QWW is important and it should be
   taken  into consideration . It is also found there is enhancement in this binding energy
as the temperature is lowered for a certain QWW radius.

\newpage
\begin{figure}
\begin{center}
\newcommand{\mathsym}[1]{{}}
\newcommand{\unicode}{{}}
\includegraphics{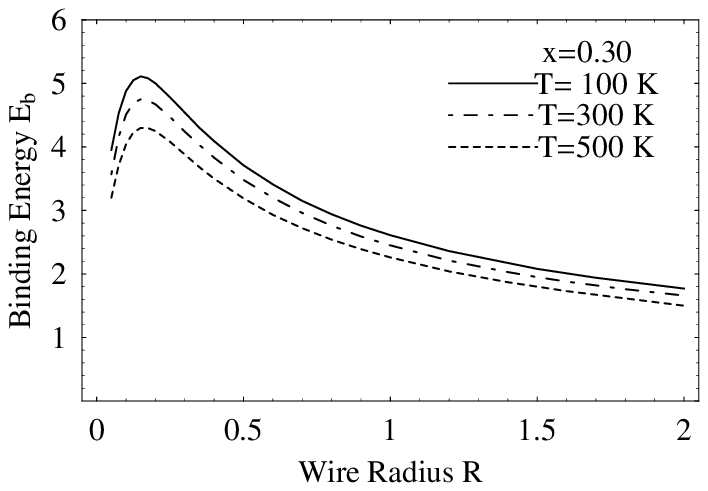}
\end{center}
\caption{Variation of the binding energy $E_b$ ,in Rydberg
$(R_b)$units as a function of the wire radius R, in Bohr radius
units ($a_B$) at x=0.30, for three Different values of
temperature, T=100 K (Solid curve), T=300 K (dashed-dotted curve),
and T=500 K (Dotted curve).}
\end{figure}

\begin{figure}
\begin{center}
\newcommand{\mathsym}[1]{{}}
\newcommand{\unicode}{{}}
\includegraphics{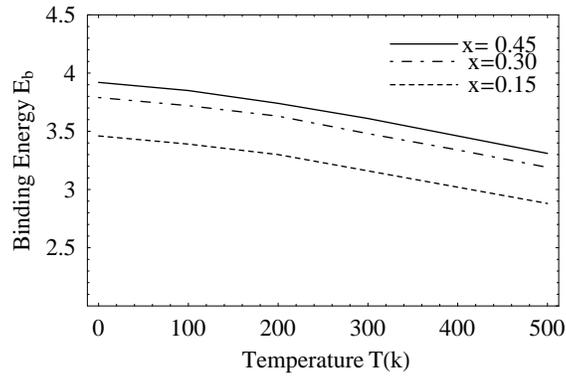}
\end{center}
\caption{Variation of the binding energy $E_b$,in Rydberg
$(R_b)$units as a function of the temperature T , in K, at a wire
radius of $1 a_B$, for three Different values of aluminium
concentration , x=0.45(Solid curve), x=0.30 (dashed-dotted curve),
and x=0.15 (Dotted curve).}
\end{figure}
\end{document}